\documentclass[%
 reprint,
superscriptaddress,
 amsmath,amssymb,
 aps,
 prl,
 floatfix
]{revtex4-2}

\usepackage{graphicx}
\usepackage{dcolumn}
\usepackage{bm}
\usepackage{hyperref}
\usepackage{xcolor}
\usepackage[T1]{fontenc} \usepackage{ae} \usepackage{aecompl}

\newcommand{\nuc}[2]{$^{#1}${#2}}

\begin{document}

\title{Magicity versus superfluidity around $^{28}$O viewed from the study of $^{30}$F}

\author{J.~Kahlbow}
\thanks{Present address: Massachusetts Institute of Technology, Cambridge, MA 02139, USA}
\email{jkahlbow@mit.edu}
\affiliation{Institut f\"ur Kernphysik, Technische Universit\"at Darmstadt, 64289 Darmstadt, Germany}
\affiliation{RIKEN Nishina Center, Hirosawa 2-1, Wako, Saitama 351-0198, Japan}

\author{T.~Aumann}
\affiliation{Institut f\"ur Kernphysik, Technische Universit\"at Darmstadt, 64289 Darmstadt, Germany}
\affiliation{GSI Helmholtzzentrum f\"ur Schwerionenforschung, 64291 Darmstadt, Germany}
\affiliation{Helmholtz Forschungsakademie Hessen f\"ur FAIR, 64291 Darmstadt, Germany}

\author{O.~Sorlin}
\affiliation{Grand Acc\'el\'erateur National d'Ions Lourds (GANIL),CEA/DRF-CNRS/IN2P3, Bvd Henri Becquerel, 14076 Caen, France}

\author{Y.~Kondo}
\affiliation{Department of Physics, Tokyo Institute of Technology, 2-12-1 O-Okayama, Meguro, Tokyo 152-8551, Japan}

\author{T.~Nakamura}
\affiliation{Department of Physics, Tokyo Institute of Technology, 2-12-1 O-Okayama, Meguro, Tokyo 152-8551, Japan}

\author{F.~Nowacki}
\affiliation{Universit\'e de Strasbourg, CNRS, IPHC UMR7178, F-67000 Strasbourg, France}

\author{A.~Revel}
\affiliation{Grand Acc\'el\'erateur National d'Ions Lourds (GANIL),CEA/DRF-CNRS/IN2P3, Bvd Henri Becquerel, 14076 Caen, France}
\affiliation{Universit\'e de Caen Normandie, ENSICAEN, CNRS/IN2P3, LPC Caen UMR6534, F-14000 Caen, France}

\author{N.~L.~Achouri}
\affiliation{Universit\'e de Caen Normandie, ENSICAEN, CNRS/IN2P3, LPC Caen UMR6534, F-14000 Caen, France}

\author{H.~Al~Falou}
\affiliation{Lebanese University, Beirut, Lebanon}

\author{L.~Atar}
\affiliation{Institut f\"ur Kernphysik, Technische Universit\"at Darmstadt, 64289 Darmstadt, Germany}

\author{H.~Baba}
\affiliation{RIKEN Nishina Center, Hirosawa 2-1, Wako, Saitama 351-0198, Japan}

\author{K.~Boretzky}
\affiliation{GSI Helmholtzzentrum f\"ur Schwerionenforschung, 64291 Darmstadt, Germany}

\author{C.~Caesar}
\affiliation{Institut f\"ur Kernphysik, Technische Universit\"at Darmstadt, 64289 Darmstadt, Germany}
\affiliation{GSI Helmholtzzentrum f\"ur Schwerionenforschung, 64291 Darmstadt, Germany}

\author{D.~Calvet}
\affiliation{Irfu, CEA, Universit\'e Paris-Saclay, 91191 Gif-sur-Yvette, France}

\author{H.~Chae}
\affiliation{IBS, 55, Expo-ro, Yuseong-gu, Daejeon, Korea, 34126}

\author{N.~Chiga}
\affiliation{RIKEN Nishina Center, Hirosawa 2-1, Wako, Saitama 351-0198, Japan}

\author{A.~Corsi}
\affiliation{Irfu, CEA, Universit\'e Paris-Saclay, 91191 Gif-sur-Yvette, France}

\author{F.~Delaunay}
\affiliation{Universit\'e de Caen Normandie, ENSICAEN, CNRS/IN2P3, LPC Caen UMR6534, F-14000 Caen, France}

\author{A.~Delbart}
\affiliation{Irfu, CEA, Universit\'e Paris-Saclay, 91191 Gif-sur-Yvette, France}

\author{Q.~Deshayes}
\affiliation{Universit\'e de Caen Normandie, ENSICAEN, CNRS/IN2P3, LPC Caen UMR6534, F-14000 Caen, France}

\author{Z.~Dombr\'adi}
\affiliation{HUN-REN Institute for Nuclear Research, HUN-REN ATOMKI, 4001 Debrecen, Hungary}

\author{C.~A.~Douma}
\affiliation{ESRIG, University of Groningen, Zernikelaan 25, 9747 AA Groningen, The Netherlands}

\author{Z.~Elekes}
\affiliation{HUN-REN Institute for Nuclear Research, HUN-REN ATOMKI, 4001 Debrecen, Hungary}

\author{I.~Ga\v{s}pari\'c}
\affiliation{Ru{\dj}er Bo\v{s}kovi\'c Institute, HR-10002 Zagreb, Croatia}
\affiliation{RIKEN Nishina Center, Hirosawa 2-1, Wako, Saitama 351-0198, Japan}

\author{J.-M.~Gheller}
\affiliation{Irfu, CEA, Universit\'e Paris-Saclay, 91191 Gif-sur-Yvette, France}

\author{J.~Gibelin}
\affiliation{Universit\'e de Caen Normandie, ENSICAEN, CNRS/IN2P3, LPC Caen UMR6534, F-14000 Caen, France}

\author{A.~Gillibert}
\affiliation{Irfu, CEA, Universit\'e Paris-Saclay, 91191 Gif-sur-Yvette, France}

\author{M.~N.~Harakeh}
\affiliation{GSI Helmholtzzentrum f\"ur Schwerionenforschung, 64291 Darmstadt, Germany}
\affiliation{ESRIG, University of Groningen, Zernikelaan 25, 9747 AA Groningen, The Netherlands}

\author{A.~Hirayama}
\affiliation{Department of Physics, Tokyo Institute of Technology, 2-12-1 O-Okayama, Meguro, Tokyo 152-8551, Japan}

\author{M.~Holl}
\affiliation{Institut f\"ur Kernphysik, Technische Universit\"at Darmstadt, 64289 Darmstadt, Germany}
\affiliation{GSI Helmholtzzentrum f\"ur Schwerionenforschung, 64291 Darmstadt, Germany}

\author{A.~Horvat}
\affiliation{Institut f\"ur Kernphysik, Technische Universit\"at Darmstadt, 64289 Darmstadt, Germany}
\affiliation{GSI Helmholtzzentrum f\"ur Schwerionenforschung, 64291 Darmstadt, Germany}

\author{\'A.~Horv\'ath}
\affiliation{E\"otv\"os Lor\'and University, P\'azm\'any P\'eter S\'et\'any 1/A, H-1117 Budapest, Hungary}

\author{J.~W.~Hwang}
\affiliation{Department of Physics and Astronomy, Seoul National University, 1 Gwanak-ro, Gwanak-gu, Seoul 08826, Republic of Korea}

\author{T.~Isobe}
\affiliation{RIKEN Nishina Center, Hirosawa 2-1, Wako, Saitama 351-0198, Japan}

\author{N.~Kalantar-Nayestanaki}
\affiliation{ESRIG, University of Groningen, Zernikelaan 25, 9747 AA Groningen, The Netherlands}

\author{S.~Kawase}
\affiliation{Department of Advanced Energy Engineering Science, Kyushu University, Kasuga, Fukuoka, 816-8580 Japan}

\author{S.~Kim}
\affiliation{Department of Physics and Astronomy, Seoul National University, 1 Gwanak-ro, Gwanak-gu, Seoul 08826, Republic of Korea}

\author{K.~Kisamori}
\affiliation{RIKEN Nishina Center, Hirosawa 2-1, Wako, Saitama 351-0198, Japan}

\author{T.~Kobayashi}
\affiliation{Department of Physics, Tohoku University, Miyagi 980-8578, Japan}

\author{D.~K\"orper}
\affiliation{GSI Helmholtzzentrum f\"ur Schwerionenforschung, 64291 Darmstadt, Germany}

\author{S.~Koyama}
\affiliation{Unversity of Tokyo, Tokyo 1130033, Japan}

\author{I.~Kuti}
\affiliation{HUN-REN Institute for Nuclear Research, HUN-REN ATOMKI, 4001 Debrecen, Hungary}

\author{V.~Lapoux}
\affiliation{Irfu, CEA, Universit\'e Paris-Saclay, 91191 Gif-sur-Yvette, France}

\author{S.~Lindberg}
\affiliation{Institutionen f\"or Fysik, Chalmers Tekniska H\"ogskola, 412 96 G\"oteborg, Sweden}

\author{F.M.~Marqu\'es}
\affiliation{Universit\'e de Caen Normandie, ENSICAEN, CNRS/IN2P3, LPC Caen UMR6534, F-14000 Caen, France}

\author{S.~Masuoka}
\affiliation{Center for Nuclear Study, University of Tokyo, 2-1 Hirosawa, Wako, Saitama 351-0198, Japan}

\author{J.~Mayer}
\affiliation{Institut f\"ur Kernphysik, Universit\"at zu K\"oln, 50937 K\"oln, Germany}

\author{K.~Miki}
\affiliation{National Superconducting Cyclotron Laboratory, Michigan State University, East Lansing, Michigan 48824, USA}

\author{T.~Murakami}
\affiliation{Department of Physics, Kyoto University, Kyoto 606-8502, Japan}

\author{M.~Najafi}
\affiliation{ESRIG, University of Groningen, Zernikelaan 25, 9747 AA Groningen, The Netherlands}

\author{K.~Nakano}
\affiliation{Department of Advanced Energy Engineering Science, Kyushu University, Kasuga, Fukuoka, 816-8580 Japan}

\author{N.~Nakatsuka}
\affiliation{Department of Physics, Kyoto University, Kyoto 606-8502, Japan}

\author{T.~Nilsson}
\affiliation{Institutionen f\"or Fysik, Chalmers Tekniska H\"ogskola, 412 96 G\"oteborg, Sweden}

\author{A.~Obertelli}
\affiliation{Irfu, CEA, Universit\'e Paris-Saclay, 91191 Gif-sur-Yvette, France}

\author{N.~A.~Orr}
\affiliation{Universit\'e de Caen Normandie, ENSICAEN, CNRS/IN2P3, LPC Caen UMR6534, F-14000 Caen, France}

\author{H.~Otsu}
\affiliation{RIKEN Nishina Center, Hirosawa 2-1, Wako, Saitama 351-0198, Japan}

\author{T.~Ozaki}
\affiliation{Department of Physics, Tokyo Institute of Technology, 2-12-1 O-Okayama, Meguro, Tokyo 152-8551, Japan}

\author{V.~Panin}
\affiliation{RIKEN Nishina Center, Hirosawa 2-1, Wako, Saitama 351-0198, Japan}

\author{S.~Paschalis}
\affiliation{Institut f\"ur Kernphysik, Technische Universit\"at Darmstadt, 64289 Darmstadt, Germany}

\author{D.~M.~Rossi}
\affiliation{Institut f\"ur Kernphysik, Technische Universit\"at Darmstadt, 64289 Darmstadt, Germany}
\affiliation{GSI Helmholtzzentrum f\"ur Schwerionenforschung, 64291 Darmstadt, Germany}

\author{A.~T.~Saito}
\affiliation{Department of Physics, Tokyo Institute of Technology, 2-12-1 O-Okayama, Meguro, Tokyo 152-8551, Japan}

\author{T.~Saito}
\affiliation{Unversity of Tokyo, Tokyo 1130033, Japan}

\author{M.~Sasano}
\affiliation{RIKEN Nishina Center, Hirosawa 2-1, Wako, Saitama 351-0198, Japan}

\author{H.~Sato}
\affiliation{RIKEN Nishina Center, Hirosawa 2-1, Wako, Saitama 351-0198, Japan}

\author{Y.~Satou}
\affiliation{Department of Physics and Astronomy, Seoul National University, 1 Gwanak-ro, Gwanak-gu, Seoul 08826, Republic of Korea}

\author{H.~Scheit}
\affiliation{Institut f\"ur Kernphysik, Technische Universit\"at Darmstadt, 64289 Darmstadt, Germany}

\author{F.~Schindler}
\affiliation{Institut f\"ur Kernphysik, Technische Universit\"at Darmstadt, 64289 Darmstadt, Germany}

\author{P.~Schrock}
\affiliation{Center for Nuclear Study, University of Tokyo, 2-1 Hirosawa, Wako, Saitama 351-0198, Japan}

\author{M.~Shikata}
\affiliation{Department of Physics, Tokyo Institute of Technology, 2-12-1 O-Okayama, Meguro, Tokyo 152-8551, Japan}

\author{K.~Shimada}
\affiliation{Department of Physics, Tokyo Institute of Technology, 2-12-1 O-Okayama, Meguro, Tokyo 152-8551, Japan}

\author{Y.~Shimizu}
\affiliation{RIKEN Nishina Center, Hirosawa 2-1, Wako, Saitama 351-0198, Japan}

\author{H.~Simon}
\affiliation{GSI Helmholtzzentrum f\"ur Schwerionenforschung, 64291 Darmstadt, Germany}

\author{D.~Sohler}
\affiliation{HUN-REN Institute for Nuclear Research, HUN-REN ATOMKI, 4001 Debrecen, Hungary}

\author{L.~Stuhl}
\affiliation{Center for Exotic Nuclear Studies, Institute for Basic Science, Daejeon 34126, Republic of Korea}
\affiliation{RIKEN Nishina Center, Hirosawa 2-1, Wako, Saitama 351-0198, Japan}

\author{S.~Takeuchi}
\affiliation{Department of Physics, Tokyo Institute of Technology, 2-12-1 O-Okayama, Meguro, Tokyo 152-8551, Japan}

\author{M.~Tanaka}
\affiliation{Department of Physics, Osaka University, Osaka 560-0043, Japan}

\author{M.~Thoennessen}
\affiliation{Facility for Rare Isotope Beams, Michigan State University, East Lansing, Michigan 48824, USA}

\author{H.~T\"ornqvist}
\affiliation{Institut f\"ur Kernphysik, Technische Universit\"at Darmstadt, 64289 Darmstadt, Germany}
\affiliation{GSI Helmholtzzentrum f\"ur Schwerionenforschung, 64291 Darmstadt, Germany}

\author{Y.~Togano}
\affiliation{Department of Physics, Tokyo Institute of Technology, 2-12-1 O-Okayama, Meguro, Tokyo 152-8551, Japan}

\author{T.~Tomai}
\affiliation{Department of Physics, Tokyo Institute of Technology, 2-12-1 O-Okayama, Meguro, Tokyo 152-8551, Japan}

\author{J.~Tscheuschner}
\affiliation{Institut f\"ur Kernphysik, Technische Universit\"at Darmstadt, 64289 Darmstadt, Germany}

\author{J.~Tsubota}
\affiliation{Department of Physics, Tokyo Institute of Technology, 2-12-1 O-Okayama, Meguro, Tokyo 152-8551, Japan}

\author{T.~Uesaka}
\affiliation{RIKEN Nishina Center, Hirosawa 2-1, Wako, Saitama 351-0198, Japan}

\author{H.~Wang}
\affiliation{RIKEN Nishina Center, Hirosawa 2-1, Wako, Saitama 351-0198, Japan}

\author{Z.~Yang}
\affiliation{RIKEN Nishina Center, Hirosawa 2-1, Wako, Saitama 351-0198, Japan}

\author{M.~Yasuda}
\affiliation{Department of Physics, Tokyo Institute of Technology, 2-12-1 O-Okayama, Meguro, Tokyo 152-8551, Japan}

\author{K.~Yoneda}
\affiliation{RIKEN Nishina Center, Hirosawa 2-1, Wako, Saitama 351-0198, Japan}

\collaboration{SAMURAI21-NeuLAND Collaboration}%
\date{July 24, 2024}%

\begin{abstract}
The neutron-rich unbound fluorine isotope $^{30}$F$_{21}$ has been observed for the first time by measuring its neutron decay at the \mbox{SAMURAI} spectrometer (\mbox{RIBF}, \mbox{RIKEN}) in the quasi-free proton knockout reaction of \nuc{31}{Ne} nuclei at $235$\,MeV/nucleon.
The mass and thus one-neutron-separation energy of \nuc{30}{F} has been determined to be $S_n = -472\pm 58 \mathrm{(stat.)} \pm 33 \mathrm{(sys.)}$\,keV from the measurement of its invariant-mass spectrum.
The absence of a sharp drop in $S_n($\nuc{30}{F}$)$ shows that the ``magic'' $N=20$ shell gap is not restored close to $^{28}$O, which is in agreement with our shell-model calculations that predict a near degeneracy between the neutron $d$ and $fp$ orbitals, with the $1p_{3/2}$ and $1p_{1/2}$ orbitals becoming more bound than the $0f_{7/2}$ one.
This degeneracy and reordering of orbitals has two potential consequences: $^{28}$O behaves like a strongly superfluid nucleus with neutron pairs scattering across shells, and both \nuc{29,31}{F} appear to be good two-neutron halo-nucleus candidates.
\newpage
\end{abstract}

\maketitle

{\emph{\textbf{Introduction}}} -- A fundamental question in nuclear physics is to understand which extreme combinations of protons and neutrons can form a bound nucleus \cite{erler12,Strob21}. 
While nucleons in stable nuclei are bound by several mega-electronvolts (MeV), adding neutrons to a given isotopic chain progressively reduces their binding energy, until reaching the so-called drip line, beyond which neutrons drip out of the nuclear potential and cannot be bound anymore. 
Irrespective of experimental efforts carried out to produce nuclei with larger neutron-to-proton imbalance, the location of the neutron drip line is so far only known up to the Ne ($Z=10$) isotopic chain~\cite{ahn19,Ahn22}. 

The study of asymmetric nuclei and their emerging phenomena serve as a critical testing ground for nuclear theories, both in terms of nuclear interactions and many-body methods~\cite{hergert20,otsuka20,nowa21}. 
The study of a given isotopic chain of light nuclei up to the proton and neutron drip lines offers the opportunity to understand nuclear structure, magic numbers, and shell evolution, as well as effects related to their weak binding energy, such as halo or cluster formation~\cite{hansen87,gnoffo22}. 
Additionally, the reduction of traditional shell gaps far from stability may lead to a transition into superfluid character in which the scattering of pairs of neutrons is enhanced. 
This superfluid phase would change from stable to neutron drip-line nuclei, i.\,e., from a regime of Cooper-like (BCS) pairs where nucleons have large correlation distances, to a Bose-Einstein Condensate (BEC)-like phase of nuclear matter, where di-neutron condensates form in a low-density environment~\cite{matsuo06,hagino07,sun10}.
The heavy fluorine isotopes ($Z=9$) discussed here are highly interesting as most of the effects mentioned above likely contribute~\cite{bagchi20, fortunato20, fossez22, singh22, luo21}.

The F isotopic chain extends up to \nuc{31}{F}, with the odd-$N$ isotopes of \nuc{28,30}{F} being unbound. 
This is as much as six neutrons further as compared to the O chain, where \nuc{24}{O} is the last bound isotope. 
There exist thus far complementary signatures for the break-down of ``magicity'' at the neutron number $N=20$ in the F isotopic chain and in the doubly-magic candidate nucleus \nuc{28}{O}. 
First, the presence of a low-lying (1/2$^+$) excited state at $1080$\,keV in \nuc{29}{F}~\cite{doornenbal17} is suggestive of the coupling of a single $d_{5/2}$ proton with a low-energy 2$^+$ excitation of the core nucleus \nuc{28}{O}. 
Second, the study of the \nuc{29}{F}$(p,pn)$ reaction~\cite{revel20} proves the dominant ground-state occupancy of the valence neutron $\ell$=1, $1p_{3/2}$ orbital, rather than the normally filled $0d_{3/2}$ orbital and the small occupancy of the $\ell$=3, $0f_{7/2}$ orbital in \nuc{29}{F}. 
This demonstrates the erosion of the $N=20$ gap between the $0d_{3/2}$ and $fp$ orbitals and an inversion between the $f$ and $p$ orbitals. 
Third, the observed increase of reaction cross section in \nuc{29}{F}, as compared to \nuc{27}{F}, is compatible with a substantial amount of intruder $p$ states~\cite{bagchi20} as found in a microscopic model that agrees with experiment and predicts an occupancy of the $0f_{7/2}$ and $1p_{3/2}$ orbits by $2.19$ and $1.26$, respectively, for \nuc{29}{F}.
Fourth, the combined facts that $S_{2n}($\nuc{29}{F}$) = 1130(540)$\,keV~\cite{gaudefroy12,wang21} is small and \nuc{31}{F} is bound imply a rather weak decrease of the two-neutron separation energy $S_{2n}$ after having passed $N=20$~\cite{gaudefroy12}. 
Fifth, the measured \nuc{29}{F}$(p,2p$)\nuc{28}{O} cross section is, when compared to theory, suggestive of a similar structure between \nuc{29}{F} and \nuc{28}{O}~\cite{kondo23}, meaning that the $N=20$ shell is not closed in \nuc{28}{O} either.

The present experimental work aims at studying for the first time the spectroscopy and the neutron separation energy $S_n$ of $^{30}$F$_{21}$, which lies one neutron beyond $N=20$. 
The combined experimental results for the F and O chains in comparison to state-of-the-art shell model calculations will introduce the discussion related to the vanishing $N=20$ gap, superfluidity, and possible halo appearance in this region of the chart of nuclides.

Technically, as for \nuc{28}{F}~\cite{revel20}, we make use of the fact that $^{30}$F is unbound to deduce its spectroscopy and neutron separation energy from the reconstructed $^{29}$F$+n$ invariant-mass spectrum, produced in $^{31}$Ne$(p,2p)^{30}$F quasi-free proton knockout reactions.
The incident \nuc{31}{Ne} displays features characteristic of a deformed nucleus with a $3/2^-$ ground state having a significant $p$-wave halo component~\cite{nakamura09,takechi12,nakamura14,chrisman21}. 
Consequently, negative parity states are expected to be populated in \nuc{30}{F} with this reaction, as protons are likely removed from the $sd$ orbitals.

{\emph{\textbf{Experimental procedure}}} --  A secondary beam of $^{31}$Ne ($1.7$\,particles/s, $235$\,MeV/nucleon) was produced at the Radioactive Isotope Beam Factory (\mbox{RIBF}) of the \mbox{RIKEN} Nishina Center by fragmentation of a $345$\,MeV/nucleon $^{48}$Ca beam ($\sim600$\,pnA) on a $15$\,mm-thick Be target, and selected using the \mbox{BigRIPS} fragment separator~\cite{kubo07}.
The $^{31}$Ne nuclei were identified via their energy loss and time-of-flight (ToF) using thin plastic scintillators.

The $^{31}$Ne secondary beam was tracked towards the \mbox{SAMURAI} superconducting dipole-magnet setup~\cite{kobayashi13} to perform the invariant-mass measurement of \nuc{30}{F} following the proton knockout reaction in the \mbox{MINOS} liquid hydrogen target of $1.14$\,g/cm$^2$~\cite{obertelli14}, using the same detector configuration as in Refs.\cite{revel20,kondo23}.
The \mbox{MINOS} target cell was surrounded by a time-projection chamber (TPC), which allowed for the reconstruction of the $(p,2p)$ reaction vertex with a precision of $3.5$\,mm ($\sigma_z$) in the beam direction. 
The DALI2 detector array~\cite{takeuchi14} placed around the target, composed of $120$ NaI(Tl) crystals, detects in-flight $\gamma$ rays with an efficiency of $\varepsilon_\gamma=13.8\%$ at $1.173$\,MeV.
The \nuc{29}{F} residues were detected at the dispersive focal plane of the \mbox{SAMURAI} spectrometer and identified by means of their ToF and energy loss information from a segmented scintillator hodoscope. 
Their momentum was obtained with a resolution of $p/\Delta p \approx 625$ ($1\sigma$) based on their bending in the \mbox{SAMURAI} magnet. 

Beam-like decay neutrons were detected at forward angles using the segmented plastic-scintillator detectors \mbox{NeuLAND} demonstrator ($400$ modules)~\cite{boretzky21} and \mbox{NEBULA} ($2\times60$ modules)~\cite{kobayashi13,nakamura16,kondo20}. 
These three walls with $40$\,cm, $24$\,cm, and $24$\,cm thickness, each accompanied by charged-particle veto detectors in front and separated by $\sim 2$\,m and $1$\,m, were placed downstream of the magnet.
The response of the neutron detectors was evaluated in a dedicated experiment using a quasi-monoenergetic neutron beam of $251$\,MeV, produced in the charge-exchange reaction \nuc{7}{Li}$(p,n)$\nuc{7}{Be}(g.s.$+430$\,keV). 
A one-neutron detection efficiency of $27.4(10)\%$ is obtained with the present \mbox{NeuLAND} demonstrator configuration, which agrees within $1\%$ with simulation. 
The overall one-neutron detection efficiency of the total array amounts to $\varepsilon_n = 54$\% for decay energies below $1$\,MeV, and decreases to $30$\% at $3.5$\,MeV.

\begin{figure}
\includegraphics[width=\columnwidth]{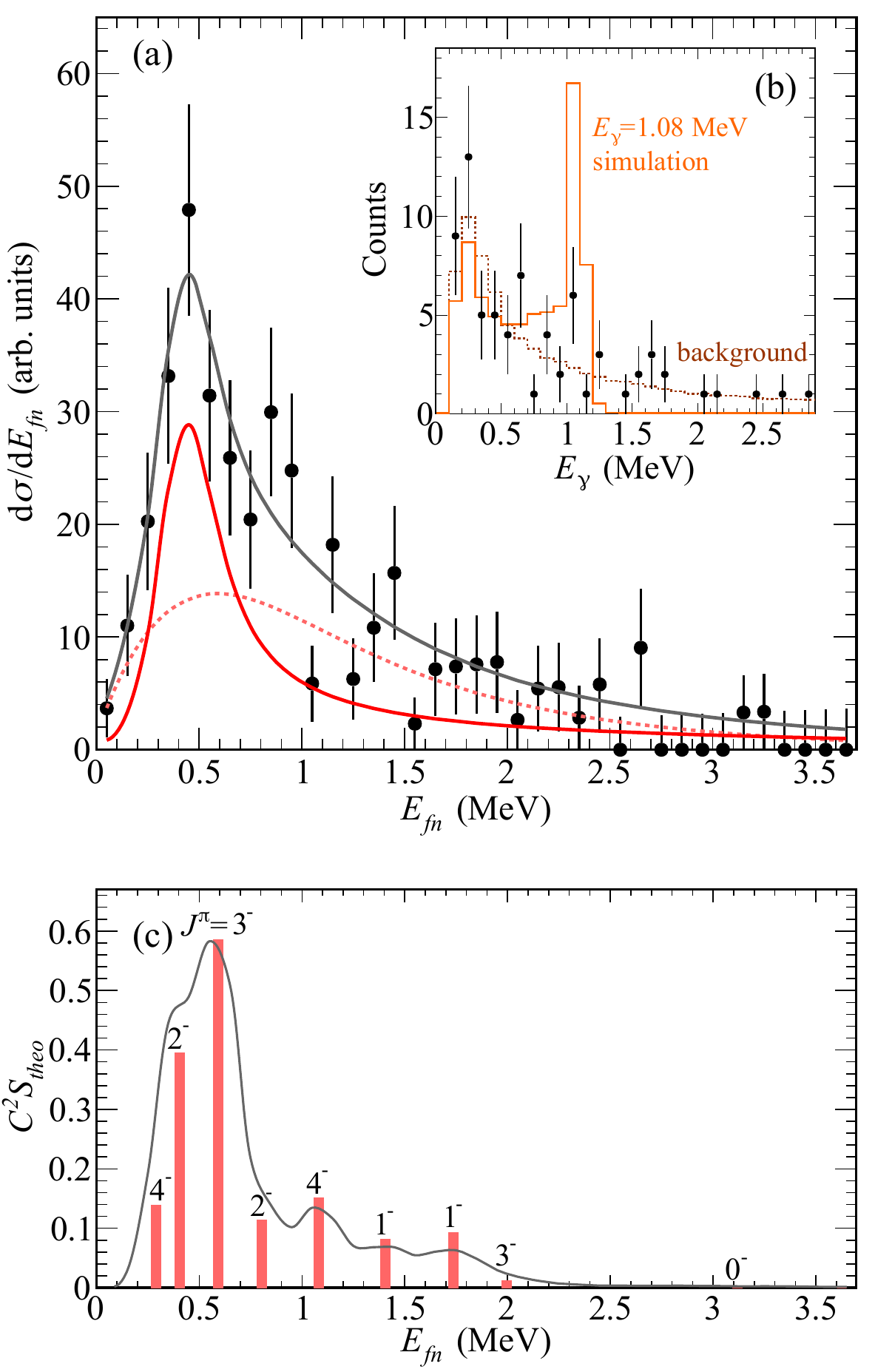}
\caption{\label{fig:erel} (a) Relative-energy spectrum of \nuc{30}{F} reconstructed
in the \nuc{31}{Ne}($p,2p$)\nuc{29}{F}$+n$ reaction. 
The data (points with $1\sigma$ stat. uncertainty) are corrected for efficiency and acceptance of the neutron detection. 
The full red curve depicts a fit with one resonance at $E^r_{fn} = 472\pm 58 \mathrm{(stat.)} \pm 33 \mathrm{(sys.)}$\,keV, while the dashed line describes unresolved resonant contributions. 
The overall gray curve shows the total fit. 
The inset (b) shows the neutron-gated Doppler-corrected $\gamma$-ray spectrum of \nuc{29}{F} in comparison to a simulated $100$\% direct $\gamma$ decay to the known $1080$\,keV state in \nuc{29}{F} (orange line) and to pure background (dashed line), extracted from the reaction \nuc{29}{F}$(p,2p)$\nuc{24}{O} in which no $\gamma$ ray is present.
The good agreement between the experimental spectrum and background proves that the observed neutron decay occurs to the g.s. of \nuc{29}{F}. 
(c) Shell-model predicted spectroscopic strength $C^2S$ of \nuc{30}{F} states produced by proton knockout from \nuc{31}{Ne} with spin-parity assignment $J^{\pi}$. 
The states have been shifted by $-600$\,keV to ease the comparison to the experimental spectrum (a). 
The gray curve depicts the summed energy spectrum in which calculated states are smeared by the experimental response.}
\end{figure}

{\emph{\textbf{Analysis \& results}}} -- \nuc{30}{F} is produced in the quasi-free \nuc{31}{Ne}$(p,2p)$ proton-knockout reaction, in which the scattered protons show coplanar kinematics with an opening angle of about $77^{\circ}$, as measured with the \mbox{MINOS} TPC. 
Being unbound, the \nuc{30}{F} nucleus immediately decays into \nuc{29}{F} and one neutron, potentially followed by $\gamma$-ray emission if \nuc{29}{F} is populated in an excited state. 
As shown in Fig.~\ref{fig:erel}(b) (and Supplemental Material~\cite{supp} Fig. S1) for the Doppler-corrected $\gamma$-ray energy spectrum in coincidence with \nuc{29}{F}$+n$, this hypothesis has been rejected based on the non-observation of a $\gamma$-ray peak, specifically at $1080(18)$\,keV which corresponds to the decay of the only excited state reported so far in \nuc{29}{F}~\cite{doornenbal17}. 
The emitted neutron is registered as the first hit in ToF in one of the neutron detectors, in coincidence with the incoming \nuc{31}{Ne} nucleus, two scattered protons, and the ($Z-1, A-2$) fragment \nuc{29}{F}.
The momentum of each decay neutron was derived from its ToF between the reaction vertex in the target and the hit position and time measured by the \mbox{NeuLAND} demonstrator or \mbox{NEBULA}~\cite{kondo20}.

Having identified the \nuc{31}{Ne}$(p,2p)$\nuc{29}{F}$+n$ reaction channel, the two-body relative energy $E_{fn}$ of the decaying \nuc{30}{F} system, shown in Fig.~\ref{fig:erel}(a), is reconstructed based on the invariant mass using the four-momenta of fragment and neutron of the two reaction products.
The experimental data points shown are corrected for the energy-dependent neutron-detection efficiency and acceptance.

The experimental $E_{fn}$ spectrum shows one clear peak. However, being an odd-odd nucleus, the level density is expected to be relatively large with, in particular, many negative-parity states as for \nuc{28}{F} (see discussion part). 
Thus, the relative-energy spectrum is likely to include unresolved resonances, the energies and widths of which cannot be constrained from the present statistics and energy resolution. 
To account for this, we fit the spectrum in a combination of a single resonance peak and a broad component including unresolved resonant contributions.
The resonance is described using a single-level energy-dependent Breit-Wigner line shape following Refs.~\cite{lane58, vandebrouck17,supp}, while the broad unresolved contribution is modeled following Ref.~\cite{caesar13} in the form of Eq.~$2$.
The line shape, folded with the experimental response matrix, and the unresolved contributions are fitted simultaneously in a $\chi^2$ minimization to the experimental data by varying the resonance energy and its width. 
The response matrix was obtained in a Geant4 simulation and relates the true and reconstructed relative energy and shape, including detector-response effects. 
The relative-energy resolution is as low as $\sim 0.10$\,MeV ($1\sigma$) at $E_{fn}=1$\,MeV and $\sim0.03$\,MeV at $0.1$\,MeV, largely driven by the performance of the more granular \mbox{NeuLAND} demonstrator.

The best fit, shown by the gray curve in Fig.~\ref{fig:erel}(a), has a reduced $\chi^2_{\mathrm{red}}$ of $0.97$. 
It results in a resonance energy of $E^r_{fn}=472\pm 58 \mathrm{(stat.)} \pm 33 \mathrm{(sys.)}$\,keV. 
The systematic uncertainty is extracted from the maximum difference for two different fit scenarios, the one described above and the other assuming a single Breit-Wigner line shape, while being largely insensitive to the choice of $\ell$ (here $\ell=1$).
While the peak value is well described and stable under various fit conditions, the extracted resonance width of $\Gamma^r=477^{+358}_{-177} \mathrm{(stat.)}$\,keV is not, because of the weakly-constrained unresolved contributions, in extreme cases it could be solely dominated by detector resolution or be a single resonance. 
Note that the quoted uncertainty has been extracted assuming a fixed contribution for the unresolved resonances. 
Assuming that no other resonance exists below this energy (see discussion below), a negative neutron separation energy of $S_n(^{30}$F$)= -472\pm 58 \mathrm{(stat.)} \pm 33 \mathrm{(sys.)}$\,keV is deduced.

{\emph{\textbf{Discussion}}} -- The following discussion is based on the experimental observation confronted with shell-model calculations. 
The calculations have been performed in the full $sd$-$pf$ valence space for neutrons and in the $sd$ shell for protons while using an updated version of the SDPF-U-MIX effective interaction, named hereafter SDPF-U-MIX20~\cite{revel20}. 
With respect to SDPF-U-MIX, monopole constraints have been incorporated in order to reproduce the $3/2^-$ and $7/2^-$ states along the $N=17$ isotones in $^{27}$Ne, $^{29}$Mg, and $^{31}$Si, as well as to fix the Duflo-Zuker mass-formula master term (Eq.~2 of \cite{MENDOZATEMIS201014}) on the binding energies of P isotopes, including those beyond $N=20$ when the $0f_{7/2}$ orbital is getting filled. 
The SDPF-U-MIX20 allows for a very good description of nuclear properties for a vast area of nuclei, ranging from $sd$ nuclei to $pf$ nuclei, including the ``island of inversion'' around \nuc{32}{Mg} and the properties of O and F at the neutron drip line. 

{\emph{Spectroscopy of $^{30}\mathrm{F}$}} -- Only one resonance could be firmly established experimentally with a width of several-hundred keV in Fig.~\ref{fig:erel}(a). 
However, our shell-model calculation predicts a large population of negative parity states in $^{30}$F (full spectrum Fig.~S3 in~\cite{supp}) through the removal of a $0d_{5/2}$ or $1s_{1/2}$ proton from the intruder neutron $3/2^-$ ground state of \nuc{31}{Ne}~\cite{nakamura09,takechi12,nakamura14}. 
The summed theoretical energy spectrum shown in Fig.~\ref{fig:erel}(c), in which each predicted state is scaled according to its spectroscopic strength $C^2S$ and smeared by the experimental response, shows good agreement with the experimental one of Fig.~\ref{fig:erel}(a) in terms of their shape, while being shifted by $-600$\,keV. 
This {\it a posteriori} supports the use of a broad contribution in the experimental fit, which would mimic several unresolved resonances with small $C^2S$ each. 
If a resonance with low $C^2S$ value existed below the experimentally found one, such as the predicted $4^-$ state, this would lead to a smaller $|S_n|$ value by about $150$\,keV. But such an hypothetical shift in $|S_n|$ would further strengthen our conclusions discussed in the following.

\begin{figure}
\includegraphics[width=\columnwidth]{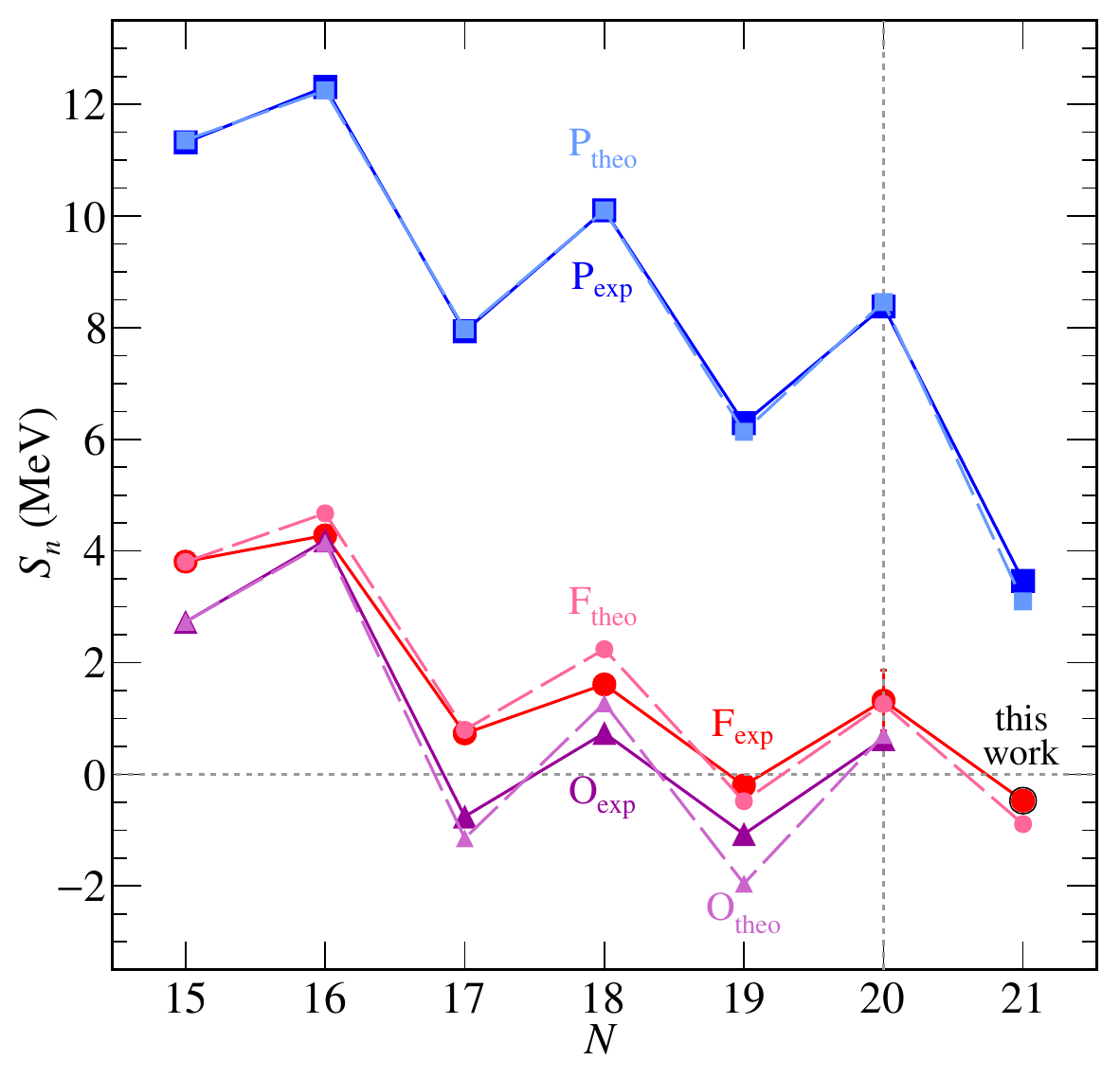}
\caption{\label{fig:sn} Experimental and theoretical neutron separation energy $S_n$ as a function of neutron number $N$ for the  fluorine F, $Z=9$ (red circles), oxygen O, $Z=8$ (magenta triangles), and phosphorus P, $Z=15$ (blue squares), isotopes. Data  points, including the new one at $N=21$ for \nuc{30}{F}, are shown with their corresponding uncertainties ($1\sigma$), often smaller than the size of the filled symbols. The theoretical results for P, F, and O, obtained from shell-model calculations using the SDPF-U-MIX20 interaction, are shown with the same symbols as data but with lighter color and linked by dashed lines instead of full lines. A compilation of $S_n$ values with more even and odd nuclei showing the damping of oscillations and the progressive vanishing of the $N=20$ shell closure towards the drip line can be found in~\cite{supp} Fig.~S4.}
\end{figure}

{\emph{Pairing and $N=20$ gap}} -- The systematics of $S_n$ values as a function of neutron number $N$ is shown for the phosphorus P, fluorine F, and oxygen O isotopes in Fig.~\ref{fig:sn}. 
In the P chain, all the experimental values are taken from Ref.~\cite{wang21}, which are well determined with less than $2$-keV uncertainty. 
The odd-even staggering along the isotopic chain is due to the presence of shell gaps at $N=16$ (mostly seen in O chain, as discussed in \cite{Lala23}) and at $N=20$, as well as pairing effects between these gaps. 
An $N=20$ gap of $4.915(13)$\,MeV can be estimated from the $S_n (^{35}$P$_{20}) - S_n (^{36}$P$_{21})$ difference.

In the F chain, the experimental $S_n$ values shown in Fig.~\ref{fig:sn} are, except for \nuc{30}{F}, taken from the atomic mass evaluation~\cite{wang21}, which makes use of the latest measurements of $^{27,28}$F from Ref.~\cite{revel20}. 
The present value of $S_n(^{30}$F$)=-472$\,keV is added to the figure. 
Even if not altering the following discussions, it is noted that the $S_n$ value of $^{26}$F could be shifted upwards by about $200$\,keV as $^{26}$F has a reported long-lived isomer state (see discussion in Ref.~\cite{lepailleur13}). 
The amplitudes of $S_n$ oscillations are similar to those in the O chain derived for $^{24-28}$O from Refs.~\cite{kondo16,kondo23}. 
However, they are much weaker in the weakly bound or unbound F than in the P chain. 
This is at variance with the empirical variation of the pairing gap following $A^{-1/2}$ and with the prediction of an increased pairing gap at low density in infinite matter~\cite{Lomb00}.

Noticeably, these oscillations remain constant in the F chain, even after passing $N=20$. This is in stark contrast to the P and other chains at $Z\gtrsim15$, in which a marked closed shell effect has been observed through a large drop in $S_n(N = 21)$ value. The absence of a sharp drop in $S_n$ value after $N=20$ in the F chain is a decisive proof that magicity is not restored close to the $Z=8,N=20$, $^{28}$O nucleus. 
A compilation of $S_n$ values towards the drip line for $Z=8$-$15$ can be found in~\cite{supp} Fig.~S4.

There is extremely good agreement between our large-scale shell-model calculations and experimental $S_n$ values in the P chain.
The agreement is also quite good for the F chain and for the known $^{24-28}$O isotopes. 
However, the calculated $S_n$ value of $^{30}$F is $-892$\,keV, which is $420$\,keV lower than the experimental one of $-472$\,keV. 
Moreover, one observes that experimental oscillations are further damped in the F and O chains by about $20$\% as compared to theory for the most neutron-rich cases. 

In the valley of stability ($Z\geq14$), a large $N=20$ gap of about $7$\,MeV is calculated using effective single-particle energies between the $0d_{3/2}$ and $0f_{7/2}$ orbitals and a smaller $N=28$ gap of about $2$\,MeV separates the $0f_{7/2}$ and the ($1p_{3/2}$, $1p_{1/2}$) orbitals (see ESPE in~\cite{supp} Fig.~S2).  
In the O and F chains, the $N=20$ gap is only $2$\,MeV and the two $1p$ orbits become more bound than the $0f_{7/2}$ one. 
The $pf$ shell inversion is also reflected by the neutron occupancy of the $0f_{7/2}$ and $1p_{3/2}$ shells of $0.2$ and $0.8$, respectively in $^{29}$F, and $0.8$ and $2.1$, respectively in $^{31}$F.
Low-$\ell$ weakly or unbound orbitals are expected to be rather sensitive to the proximity of the continuum~\cite{Hamm01, Kay17} which may further reduce the $N=20$ gap. 
Theory frameworks of Refs.~\cite{michel20,fossez22} that incorporate continuum degrees of freedom reproduce well the constancy and damping of $S_n$ oscillations in the F chain even after having passed $N=20$ (cf. Fig.~S5 of~\cite{supp}), in particular the Gamow shell model calculation using the Furutani-Horiuchi-Tamagaki interaction~\cite{michel20}.

{\emph{Superfluidity}} -- The close proximity of all neutron orbits and the fact that the proton $Z=8$ core is likely preserved induce interesting properties for the collectivity (quadrupole, pairing) of these nuclei. 
According to our calculations, $^{28}$O has $97$\% of pairs coupled to $J=0$ (seniority $0$), with $50$\% closed-shell configuration and $47$\% of pairs involved in $sd$ to $fp$ excitations, mostly between the $d_{3/2}$ and the $p_{3/2}$ orbitals. 
The $^{29}$F nucleus is also dominated by pairs coupled to $J=0$ (by $70$\%), with more than $60$\% of scatter from the $sd$ to $pf$ orbits. 

Such a regime in which pairs of nucleons equivalently occupy nearby orbitals is usually defined as superfluid (see e.\,g. Ch. $6$ of Ref.~\cite{RingSchuck}). 
A characteristic example of superfluidity is found in the Tin isotopic chain where, between $N=50$ and $N=82$, ($J=0$)-coupled neutron pairs scatter between all nearby orbitals, keeping the $Z=50$ proton core mainly unaffected at low excitation energy. 
This has the consequence of observing almost constant pairing oscillations in the whole Tin isotopic chain. 
The constancy of $S_n$ oscillations in the F chain, experimentally established even above $N=20$, and predicted to continue in $^{31}$F$_{22}$ with an $S_n$ value of $924$\,keV, is a likely consequence of the mixing between orbitals and a further indication of the onset of this superfluid regime.
The reduced amplitude of oscillations in the O and F chains, as compared to the P chain for instance, can partially be explained by the mixed filling of the $d_{3/2}$ and $p_{3/2}$ orbitals.
Indeed, the amplitude of the odd-even pairing oscillations are weaker in the $p_{3/2}$ orbital than in the $d_{3/2}$, as can be deduced from the observed amplitudes of pairing oscillations in the Ca isotopic chain with the sequential filling of the $d_{3/2}, f_{7/2}, p_{3/2}$, and $p_{1/2}$ orbitals~\cite{Lala23}.
A similar reduction of $S_n$ oscillations has been found in the neutron-rich B isotopes~\cite{gaudefroy12,leblond18}.

As for the  nature of the superfluid phase in the (O,F) isotopic chains, the mixing between weakly bound orbits of different parities should favor the transition to a BEC regime with neutron pairs of much smaller size as compared to BCS according to Refs.~\cite{Cata84,fortunato20,Ober21,Yang23}. 
This transition is planned to be explored on a theoretical ground using the relative coordinate of neutrons in shell-model calculations. 
Future experimental works should focus on the determination of the relative distance or angle between neutrons using experimental techniques such as the ones described in Refs.~\cite{Naka06,Kubo20}.

{\emph{Two-neutron halos}} -- As discussed in Refs.~\cite{bagchi20,michel20,fortunato20,singh22} the presence of these low-$\ell$ orbitals, their high occupancy, and their relatively weak binding likely favors the development of two-neutron halo structures in $^{29}$F and more importantly in $^{31}$F.
This is corroborated by our shell-model calculations based on the large occupancy of the $p_{3/2}$ orbital by $0.8$ and $2.1$ neutrons for $^{29}$F and $^{31}$F, respectively, and our predicted low $S_{2n}$ values of $776$\,keV for $^{29}$F, to be compared to the experimental value of $1130(540)$\,keV~\cite{gaudefroy12,wang21}, and $32$\,keV for $^{31}$F.

{\emph{\textbf{Summary}}} -- In this letter, we report on the first study of the neutron-rich \nuc{30}{F} nucleus, produced by means of quasi-free proton knockout reactions of a \nuc{31}{Ne} beam on a liquid hydrogen target. 
The excitation energy spectrum of \nuc{30}{F} and its neutron separation energy $S_n$ have been obtained using the invariant-mass method from the decay into \nuc{29}{F}$ + n$. 
Based on the deduced $S_n($\nuc{30}{F}$)$ value of $-472\pm 58 \mathrm{(stat.)} \pm 33 \mathrm{(sys.)}$keV, we find no sharp decrease in $S_n$ after $N=20$ but rather damped pairing oscillations of constant amplitude. 
The absence of a sharp drop is the most direct confirmation that the $N=20$ shell gap is not restored in this region of the chart of nuclides close to \nuc{28}{O}.
The constancy of oscillations, as well as their weaker amplitude as compared to the P isotones are  suggestive of a  coupling to the continuum, as well as of the mixing between the nearby $0d_{3/2}$ and $1p_{3/2}$ orbitals.  This mixing induces the establishment of a superfluid regime, which is further substantiated by shell model calculations in which $^{28}$O and $^{29}$F have strongly dominating configurations of neutron pairs coupled to $J=0$, scattering by about half between the $sd$ and $fp$ shells. 
Searching for a transition to the superfluid BEC phase in the F chain and studying the two-neutron halo feature of $^{31}$F are imminent topics for the future.

\begin{acknowledgments}
{\emph{\textbf{Acknowledgments}}} -- We thank the accelerator staff of the \mbox{RIKEN} Nishina Center for their efforts in delivering the intense \nuc{48}{Ca} beam. 
This project was supported by Deutsche Forschungsgemeinschaft (DFG, German Research Foundation), Project-ID 279384907 -- SFB 1245, and the GSI-TU Darmstadt cooperation agreement.
J.K. also acknowledges support from \mbox{RIKEN} as short-term International Program Associate and from Tokyo Institute of Technology under the Foreign Graduate Student Invitation Program.
T.N. and Y.K. acknowledge support from JSPS KAKENHI Grant Nos. JP18H05404, JP21H04465, and JP24H00006. Y.K. also acknowledges partial support from JSPS KAKENHI Grant No. JP18K03672.
N.L.A., F.D., J.G., F.M.M., and N.A.O. acknowledge partial support from the Franco-Japanese LIA-International Associated Laboratory for Nuclear Structure Problems as well as the French ANR-14-CE33-0022-02 EXPAND. 
I.G. was supported by the Helmholtz International Center for FAIR and Croatian Science Foundation under Projects No. 1257 and No. 7194. 
This work was also supported in part by the National Science Foundation, USA under Grant No. PHY-1102511.
Z.D., Z.E., and D.S. were supported by National Research, Development and Innovation Fund of Hungary (NKFIH) Projects No. TKP2021-NKTA-42 and No. K147010, and I.K. by Project No. PD 124717.
L.S. acknowledges support from the Institute for Basic Science under IBS-R031-D1.
H.W. acknowledges the support by JSPS KAKENHI Grant Nos. JP21H04465 and JP18H05404.
\end{acknowledgments}

\bibliography{F30_references}%
\nocite{luo02}
\nocite{hu20}


\end{document}